\documentclass[default]{sn-jnl}

\usepackage{graphicx, caption}%
\usepackage{multirow}%
\usepackage{amsmath,amssymb,amsfonts}%
\usepackage{amsthm}%
\usepackage{mathrsfs}%
\usepackage[title]{appendix}%
\usepackage{xcolor}%
\usepackage{textcomp}%
\usepackage{manyfoot}%
\usepackage{booktabs}%
\usepackage{algorithm}%
\usepackage{algorithmicx}%
\usepackage{algpseudocode}%
\usepackage{listings}%

\usepackage{makecell}
\usepackage{tabularx}
\usepackage{arydshln}
\usepackage{siunitx}

\theoremstyle{thmstyleone}%
\theoremstyle{thmstyletwo}%

\theoremstyle{thmstylethree}%

\begin{document}

\title[Article Title]{QUEST-DMC: Background Modelling and Resulting Heat Deposit for a Superfluid Helium-3 Bolometer}


\author[1]{\fnm{S.} \sur{Autti}}
\author[2]{\fnm{A.} \sur{Casey}}
\author[2]{\fnm{N.} \sur{Eng}}
\author[2]{\fnm{N.} \sur{Darvishi}}
\author[1,2]{\fnm{P.} \sur{Franchini}}
\author[1]{\fnm{R.P.} \sur{Haley}}
\author[2]{\fnm{P.J.} \sur{Heikkinen}}
\author[3]{\fnm{A.} \sur{Kemp}}
\author*[2,3]{\fnm{E.} \sur{Leason}}\email{elizabeth.leason@rhul.ac.uk}
\author[2]{\fnm{L.V.} \sur{Levitin}}
\author[3]{\fnm{J.} \sur{Monroe}}
\author[3]{\fnm{J.} \sur{March-Russel}}
\author[1]{\fnm{M.T.} \sur{Noble}}
\author[1]{\fnm{J.R.} \sur{Prance}}
\author[2]{\fnm{X.} \sur{Rojas}}
\author[1]{\fnm{T.} \sur{Salmon}}
\author[1]{\fnm{J.} \sur{Saunders}}
\author[1]{\fnm{R.} \sur{Smith}}
\author[1]{\fnm{M.D.} \sur{Thompson}}
\author[1]{\fnm{V.} \sur{Tsepelin}}
\author[2]{\fnm{S.M.} \sur{West}}
\author[1]{\fnm{L.} \sur{Whitehead}}
\author[4]{\fnm{K.} \sur{Zhang}}
\author[1]{\fnm{D.E.} \sur{Zmeev}}

\affil[1]{\orgdiv{Department of Physics}, \orgname{Lancaster University}, \orgaddress{ \city{Lancaster} \postcode{LA1 4YB}, \country{UK}}}
\affil[2]{ \orgdiv{Department of Physics}, \orgname{Royal Holloway University of London}, \orgaddress{\city{Egham}, \postcode{TW20 0EX}, \country{UK}}}
\affil[3]{ \orgdiv{Department of Physics}, \orgname{University of Oxford}, \orgaddress{\city{Oxford}, \postcode{OX1 3PJ}, \country{UK}}}
\affil[4]{ \orgdiv{Department of Physics}, \orgname{University of Sussex}, \orgaddress{\city{Brighton}, \postcode{BN1 9QH}, \country{UK}}}


\abstract{


We report the results of radioactivity assays and heat leak calculations for a range of common cryogenic materials, considered for use in the QUEST-DMC superfluid $^3$He dark matter detector. The bolometer, instrumented with nanomechanical resonators, will be sensitive to energy deposits from dark matter interactions. Events from radioactive decays and cosmic rays constitute a significant background and must be precisely modelled, using a combination of material screening and Monte Carlo simulations. However, the results presented here are of wider interest for experiments and quantum devices sensitive to minute heat leaks and spurious events, thus we present heat leak per unit mass or surface area for every material studied. This can inform material choices for other experiments, especially if underground operation is considered -- where the radiogenic backgrounds will dominate even at shallow depths.}

\maketitle

\section{Introduction}\label{sec:intro}

The nature of dark matter remains an open question in fundamental physics, with extensive direct, indirect and collider searches all returning null results. These searches have typically focused on GeV/c$^2$ - TeV/c$^2$ mass particle dark matter. An increasing number of experiments are also investigating ultra-light boson dark matter with masses much below eV/c$^2$, using techniques to search for wavelike phenomena. However, low mass particle dark matter in the intermediate mass range is not well constrained.


Superfluid helium is an attractive target for low mass dark matter searches due to good kinematic matching, intrinsic radiopurity and small superfluid energy gap.
The HeRALD \cite{Herald_2019} and DELight \cite{Delight_2022} collaborations are investigating the use of $^4$He target for a dark matter search. The QUEST-DMC collaboration is exploring the complementary use of superfluid $^3$He to search for spin-dependent dark matter interactions in the sub-GeV mass range \cite{quest_epjc}. With a projected energy threshold of 0.51 eV for nuclear recoil interactions, we expect to be able to probe dark matter masses down to $\sim 25$ $\rm MeV/c^2$ and spin-dependent cross sections down to $\sim 10^{36}$ cm$^2$ with a 4.9 g day exposure, from a 6 month run.

\subsection{QUEST-DMC Experiment}\label{sec:QUEST}

The idea of using $^3$He as a bolometer for particle detection dates back to 1988 \cite{Pickett_1988} and was explored by the MACHe3 \cite{Mache3_2000} and ULTIMA projects \cite{Winkelmann_2007}. In the QUEST-DMC experiment the superfluid $^3$He target will be enclosed in a $\sim$ 1 cm$^3$ transparent box instrumented with a nanomechanical resonators (NEMS) \cite{nems_preprint}. This is surrounded by a secondary superfluid volume with  connection via a $\sim$ 1 mm$^2$ hole in the bolometer wall.  Energy deposition following a dark matter scattering interaction with $^3$He leads to the production of quasiparticles (broken Cooper pairs) and scintillation photons (following excitation and ionization processes). The quasiparticles are detected as a damping force on the NEMS driven on resonance. Superfluid $^3$He in the bolometer is cooled to around $\SI{100}{\micro K}$ to ensure a small thermal population of quasiparticles and maximise sensitivity of the detector to generated quasiparticles. Scintillation photons can be detected using photon sensors surrounding the bolometer. For a complete description of the detector and operation see Ref. \cite{quest_epjc}.

\section{Background Modelling}\label{sec:bgmodelling}

Energy deposition from particles such as cosmic rays or radioactive decay products interacting with target atoms can mimic a dark matter interaction --- a significant background in a rare event search. Modelling those events using Monte Carlo simulation and material screening is important for experiment design, assessment of the projected dark matter sensitivity and eventual limit setting. In the context of QUEST-DMC, detailed modelling of energy deposits in the system will also be useful for future studies of superfluid helium physics.

Background sources can be external to the experiment --- cosmic rays, neutrons and $\gamma$ rays coming from the surroundings, or internal --- radioisotope decays in the detector materials, surfaces or the target itself. The incoming particles can transfer energy to the target through interactions with either electrons or nucleons, which result in the production of quasiparticles and photons described above. External backgrounds depend on the experiment location and can be mitigated using shielding or external veto tagging  detectors. Radiogenic backgrounds from detector materials can be minimised using careful design choices in material selection and detector geometry. Internal radiogenic backgrounds from intrinsic contaminants can be minimised by improving material purity and surface contaminants can be minimised using strict cleaning protocols. Once mitigations are implemented it is important to accurately understand the expected background levels, in order to establish or rule out the presence of any candidate dark matter signal.

Superfluid $^3$He is intrinsically radiopure --- at this operating temperature impurities will have frozen out before entering the bolometer. The only other possible isotope is $^4$He, but the low solubility and preferential adsorption on the cell walls mean that no $^4$He atoms are expected in the bulk liquid at sub-millikelvin temperatures. Therefore, our assessment of radiogenic backgrounds will focus on naturally occurring radioisotopes embedded in the detector materials. The most common radioisotopes are: $\rm ^{238}$U, $\rm ^{235}$U, $^{232}$Th, $^{40}$K, $^{60}$Co and $^{137}$Cs. Uranium and thorium isotopes and their progeny form chains which decay through multiple $\alpha$ and $\beta$ (and subsequent $\gamma$) emissions to eventually form stable lead isotopes. The isotopes $^{40}$K, $^{60}$Co and $^{137}$Cs undergo single $\beta$ decays, with subsequent $\gamma$ ray emission.

The  $\rm ^{238}$U and $^{232}$Th decay chains are typically assumed to be in secular equilibrium, due to the long lifetimes of parent nuclei relative to their daughters, allowing a measurement of activity in one part of the chain to determine the activity of the rest. However, secular equilibrium can be broken in both chains by enrichment or removal of radium. In the $^{238}$U chain there is a simple equilibrium break at $^{226}$Ra, which has a half life of 1600 years so any change will take thousands of years to be restored. The chain can be divided into ``early" --- for isotopes above $^{226}$Ra and ``late" --- for $^{226}$Ra and below, both of which are in secular equilibrium. Similarly, the $^{232}$Th chain is split into ``early", above $^{224}$Ra, and ``late", including and below $^{224}$Ra.

\subsection{Material Screening}\label{subsec:screening}

A range of spectroscopic assay techniques are used to measure radioisotope activity from materials, with each method sensitive to different radiation types and energy ranges. Commonly used measurements include high purity germanium (HPGe) spectrometry, inductively coupled plasma mass spectrometry (ICP-MS), radon emanation and alpha detection. For the uranium and thorium chains at keV-MeV energies HPGe is the most relevant technique and has the advantage of being non-destructive. HPGe assays use a Ge crystal for gamma spectroscopy, to determine levels of naturally occurring radioisotopes in a sample through detection of $\gamma$ rays associated with their decay. This technique cannot distinguish between decays happening on the material surface or in the bulk.

The Boulby UnderGround Screening (BUGS) facility, located 1.1 km underground in Boulby Mine, was used to perform HPGe measurements of materials for the QUEST-DMC experiment. The BUGS facility, originally dedicated to HPGe, contains seven HPGe detectors in a class 1000 cleanroom, with ICP-MS, alpha detection and radon emanation facilities added later \cite{Scovell_2018, Scovell_2023}. For these measurements the ultra-low background detectors Chaloner and Lunehead and speciality ultra-low background detector Roseberry were used. The detectors are housed inside multi-layer castles, consisting of 10 cm high-purity copper and 10 cm lead, to shield them from environmental $\gamma$ rays. Detector materials used inside the castle are specially selected based on low radioactivity and the manufacturing process is carefully controlled to minimise contaminants. The castle is purged using N$_2$ gas to remove airborne radon, with residual radon in the N$_2$ removed using charcoal traps.

Detectors in the BUGS facility have a range of different types and configurations to allow for a range of different sample geometries and cover a large range of $\gamma$ ray energies. Roseberry is a Mirion BE6530 Broad Energy Germanium (BEGe) planar detector with a 65 cm$^2$ face and 30 mm thickness, giving high sensivity to low energy $\gamma$ rays. Chaloner is a Mirion BE5030 BEGe planar detector with a 50 cm$^2$ face and 30 mm thickness, again giving high efficiency for low energy gammas, but a small volume (150 cm$^2$) best suited for small samples. Lunehead is a p-type coaxial Ortec GEM-XX240-S detector, with a larger (370 cm$^2$) volume but reduced sensitivity to the lowest energy $\gamma$ rays. The 46.5 keV $\gamma$ ray emission from $^{210}$Pb cannot be detected by Lunehead, preventing measurement of the $^{210}$Pb activity.

\begin{sidewaystable}
\centering\setcellgapes{2pt}\makegapedcells
\caption{Samples and measured activities for gamma spectroscopy carried out by the Boulby Underground Germanium Screening facility. Where upper limits are reported these are at the 90\% confidence level. The Lunehead detector has no sensitivity to  $^{210}$Pb activity. Two further results not included in the table, in units mBq/kg: Silver sinters Ag$^{108m}$ $42 \pm 5$ and Painted Al $^{137}$Cs $5.4 \pm 0.8$. For comparison two samples have been added from the SNOLAB radiopurity database \cite{radiopurity} - OFHC Cu screened by XENON1T \cite{XENON1T_radioassay} and kapton Cu PCBs screened by TREX \cite{TREX_radioassay} - note that reported upper limits for these samples are at the 95\% confidence level. Results not shown in the table for these samples, in units mBq/kg: Cu $^{235}$U $<1.8$, $^{60}$Co $0.07 \pm 0.01$, $^{137}$Cs $< 0.008$, Kapton Cu PCBs $^{235}$U $<0.12$, $^{60}$Co $<1.06$, $^{137}$Cs $<1.23$.} \label{tab:screening}
\begin{tabular*}{\linewidth}{lcccccccccc}
\\\cmidrule{0-8}%
& & & \multicolumn{6}{c}{Measured activity [mBq/kg] } \\\cmidrule{4-9}%
Sample & Mass [g] & Detector & $\rm ^{238} U_{early}$ & $\rm ^{238}U_{late}$ &  $\rm ^{210}Pb$ & $\rm ^{232}Th_{early}$ & $\rm ^{232}Th_{late}$ & $\rm ^{40}K$ \\
\\\cmidrule{0-8}%
Stainless steel &  544.2 & Roseberry & $16(8)$ & $2.5(0.9)$ & $82(27)$ & $3.1(1.2)$ & $3.9(0.9)$ & $<6.2$\\
Al 6061-O &  642.6 & Lunehead & $8330(270)$ & $15.3(3.9)$ & - & $356(12)$ & $334.4(8.2)$ & $56(8)$ \\
Painted Al &  923.0 & Chaloner & $25680(230)$ & $16.2(3.2)$ & $60480(540)$ & $259.2(8.3)$ & $342.2(6.2)$ & $21.8(9.6)$ \\
Brass &  107.0 & Roseberry & $<7.6$ & $4(1)$ & $14990(350)$ & $<1$ & $<1.1$ & $<7.3$ \\
Silver sinters &  37.1 & Roseberry & $<90$ & $<36$ & $430(320)$ & $<27$ & $<28$ & $<385$  \\
Vespel & 38.3 & Chaloner & $87 \pm 66$ & $90(14)$ & $418 \pm 85$ & $111(25)$ & $64(14)$ & $430(240)$\\
Fiberglass &  6.02 & Chaloner & $32580(640)$ & $15154(62)$ & $68600(1400)$ & $11400(100)$ & $12005(62)$ & $23520(440)$ \\
Araldite &  161.9 & Roseberry & $< 3.6$ & $<4.8$ & $14.5(9.7)$ & $<3.4$ & $<2.2$ & $<25.5$ \\
Stycast &  131.5 & Chaloner & $<10.5$ & $<9.5$ & $<14.9$ & $<12.9$ & $<6.2$ & $<122.2$ \\
GRP &  106.9 & Lunehead & $5700(1000)$ & $7460(120)$ & - & $7840(160)$ & $7350(100)$ & $4900(570)$ \\
PEN &  35.1 & Roseberry & $<4.2$ & $6.4(2.7)$ & $26(13)$ & $<3.4$ & $<2.4$ & $<22.8$
\\
\hdashline 
\\[-2.0ex]
OFHC Cu & 8800.0 & GeMPI I & $<1.9$ & $<0.13$ & - & $0.090(4)$ & $0.090(4)$ & $0.5 (2)$ \\
Kapton Cu PCB & - & Ge Paquito & $<184.9 $& $<5.7$ & - & $<4.84$ & $<2.91$ & $<17.6$ \\
\cmidrule{0-8}%
\end{tabular*}
\end{sidewaystable}

In the process of designing the QUEST-DMC experiment and evaluating the sensitivity to dark matter interactions eleven materials, commonly used in ultra-low temperature physics experiments, have been screened for 1-2 weeks at the BUGS facility. Surfaces of the samples were cleaned with lint-free wipes using isopropyl alcohol, to minimise surface contamination. Samples, detector details and measured radioisotope activity levels in these materials are shown in Table \ref{tab:screening}. The samples screened consisted of cryostat metal parts, other cryostat materials and candidate materials for the experimental cell, described below:

\begin{itemize}
\item stainless steel grade 304, used for the vacuum can -- 15 cm square sheet sample, $< 0.5$ cm thickness
\item aluminium 6061-O, with and without paint, used to make the helium dewar surrounding the experiment -- machined disks, 14-19 cm diameter  
\item brass grade CZ121, used for radiation shield cap -- machined disk, 6 cm diameter
\item silver sinters, immersed in the $^3$He for thermal coupling -- four blocks, 5 cm length
\item vespel pillars, used as thermally-insulating mechanical supports between different stages of the dilution refrigerator -- three hollow cylindrical pillars, 6 cm length
\item copper coated fiberglass PCB -- $(20 \times 4)$ cm sheet, $< 0.5$ cm thickness
\item Stycast 1266 epoxy manufactured by Henkel (parts A and B, with 100:28 mixing ratio by mass), used for experimental cell -- single cuboid $\sim (8 \times 6 \times 2.5)$ cm
\item Araldite epoxy -- two cylindrical pieces, 4 cm diameter, combined height $\sim 10$ cm
\item glass reinforced plastic (GRP) black nylon 66 with 30\% glass reinforcement, possible experimental cell material -- three cylindrical pieces, 4 cm diameter, $\sim 1$ cm height
\item polyethylene naphthalate (PEN), possible cell material or wavelength shifter -- 1 m folded sheet
\end{itemize}

In addition, we can make use of assay results previously reported by other groups, which are extensively catalogued in the SNOLAB radiopurity database \cite{radiopurity}. 
Two examples of candidate materials screened by other dark matter experiments are; oxygen-free high conductivity (OFHC) copper C10100 by the XENON1T collaboration \cite{XENON1T_radioassay} and kapton copper monolayer printed PCBs by the TREX collaboration \cite{TREX_radioassay}. The screening results for these materials, chosen specifically for their low activities, are shown in Table \ref{tab:screening} for comparison. It is also important to note that the level of contaminants in composite materials such as alloys and epoxies can vary significantly between grades or batches. For example, different brass grades can show significant differences in $^{210}$Pb activity, as demonstrated in screening results reported in Ref. \cite{EDEL_radioassay}. Stycast and Araldite epoxy variation between batches depends on the ratios and mixing process e.g. Araldite screening results reported on \cite{radiopurity} have $\rm ^{238}U_{early}$ activities varying from $22.2 \pm 2.5$ mBq/kg to $119.8 \pm 6.2$ mBq/kg.

\subsection{Heat Leak}\label{subsec:heatleak}

\begin{table}[b!]
\captionsetup{width=\linewidth}
\caption{Emitted power per unit mass for each sample, based on the screening results reported and calculation reported above.}\label{tab:heat_leak}%
\begin{tabular}{@{}llll@{}}
\toprule
& \multicolumn{3}{c}{Emitted Power [pW/kg]} \\ 
Sample & Alpha & Beta  & Gamma\\
\midrule
Unpainted Al  & $20.1(3)$ & $1.15(3)$ & $0.252(4)$ \\
Painted Al & $110.0(5)$  & $7.21(4)$  & $0.374(3)$ \\
Stainless steel  & $0.14(2)$  & $0.009(2)$  & $0.0036(4)$ \\
Brass & $12.8(3)$ & $1.43(2)$ & $0.022(4)$ \\
Silver sinters & $0.9(3)$  & $0.06(2)$ & $0.04(1)$ \\
Vespel & $1.4(1)$ & $0.07(1)$ & $0.082(9)$ \\
Fiberglass & $262(1)$  & $1.5(1)$ & $12.51(3)$ \\
Araldite & $0.06(1)$ & $0.0027(7)$ & $0.0037(6)$ \\
Stycast & $0.13(2)$ & $0.004(1)$ & $0.000(2)$ \\
GRP & $152(2)$ & $10.0(2)$ & $8.13(5)$ \\
PEN & $0.07(1)$ & $0.0035(9)$ & $0.006(6)$ \\
\hdashline 
\\[-2.0ex]
OFHC Cu & $0.005(2)$ & $0.0003(1)$ & $0.00015(3)$ \\
Kapton Cu PCB & $4.48(2)$ & $1.12(2)$ & $0.0049(9)$\\
\botrule
\end{tabular}
\end{table}

The measured activities from material screening can be converted into a heat leak for each material in pW per kg. This is done by considering all $\alpha$, $\beta$ and $\gamma$ emitting processes for the U, Th decay chains and individual radioisotopes. Sensitivity to $^{235}$U activity is very low in measurements with the sample masses and exposure times above. When no $^{235}$U activity could be measured the theoretical ratio of natural abundance $^{235}$U/$^{238}$U = 0.007257 is used in the heat leak calculations \cite{RSC_Uratio}. The power $P$, is calculated by taking the product of energy $E_i$, activity $a_i$, branching ratio $br_i$ for every decay in each chain then the summing over all decays of a given type:
\begin{equation}
P = \sum_i E_i \times a_i \times br_i.
\end{equation}

For the $\alpha$ and $\gamma$ emissions $E_i$ is the discrete energy of the decay, whilst for $\beta$ emission the mean energy of the emitted spectrum is used. Only decays with branching ratios greater than 1\% are considered, branching ratio and decay energies are taken from the ENSDF database \cite{ENSDF}. Resulting powers per unit mass of sample are shown in Table \ref{tab:heat_leak} for the different decay types. 

\begin{table}[t!]
\captionsetup{width=\linewidth}
\begin{tabular}{@{}lll@{}}
\toprule
Sample & Alpha power [pW/m$^{2}$] \\
\midrule
Al  & $0.85(1)$  \\
Painted Al & $5.76(3)$   \\
Stainless  & $0.015(2)$   \\
Araldite & $0.0021(5)$  \\
Stycast & $0.004(1)$  \\
GRP & $6.37(50)$  \\
Brass & $1.15(3)$  \\
Silver sinters & $0.10(3)$   \\
Vespel & $1.05(1)$  \\
Fiberglass & $19.0(5)$  \\
PEN & $0.0024(5)$ \\
\hdashline 
\\[-2.0ex]
OFHC Cu & $0.0004(1)$\\
Kapton Cu PCB & $0.016(4)$\\  
\botrule
\end{tabular}
\caption{Emitted power per unit area for $\alpha$ decays from a given sample material. These units are used since the stopping length of $\alpha$ particles is below 1 mm in most materials, so not all $\alpha$ particles emitted in the bulk will escape the surface of a material.}
\label{tab:alpha_sa}%
\end{table}

It is important to note that the stopping powers of the three radiation types will differ, so their expected ranges will vary. For example an $\alpha$ particle with kinetic energy of 1 MeV will have a range of $\sim \SI{3}{\micro\meter}$ in aluminium, whilst a 1 MeV $\beta$ will have a $\sim$ 2 mm range and 1 MeV $\gamma$ a $\sim$ 6 cm range \cite{astar}. Therefore many of the emitted $\alpha$ particles will not escape source material which is more than $\SI{}{\micro\meter}$ thickness, so we can also report the $\alpha$ heat leak per unit surface area. Tabulated mass range values (g/cm$^2$) as a function of $\alpha$ energy are taken from the ASTAR database \cite{astar} for different materials and interpolated to find the range corresponding to the decay product energy. The power corresponding to each decay is again calculated from the product of energy, activity, branching ratio and range. The sum is then taken over all decays to give the total power emitted per unit surface area, reported in Table \ref{tab:alpha_sa}.

These results allow for a comparison of relative radioactive emission by different materials, however the decay products can be attenuated or stopped by surrounding materials and the heat leak in a given volume will depend strongly on the geometry and combination of materials used. Power emitted by certain popular construction materials makes them unusable for ultra-low temperature experiments. It is careful choice of materials that ultimately defines the lowest achievable experimental temperatures.

\subsection{Background Simulations}\label{subsec2}

\begin{figure}[b!]%
\centering
\includegraphics[width=0.99\textwidth]{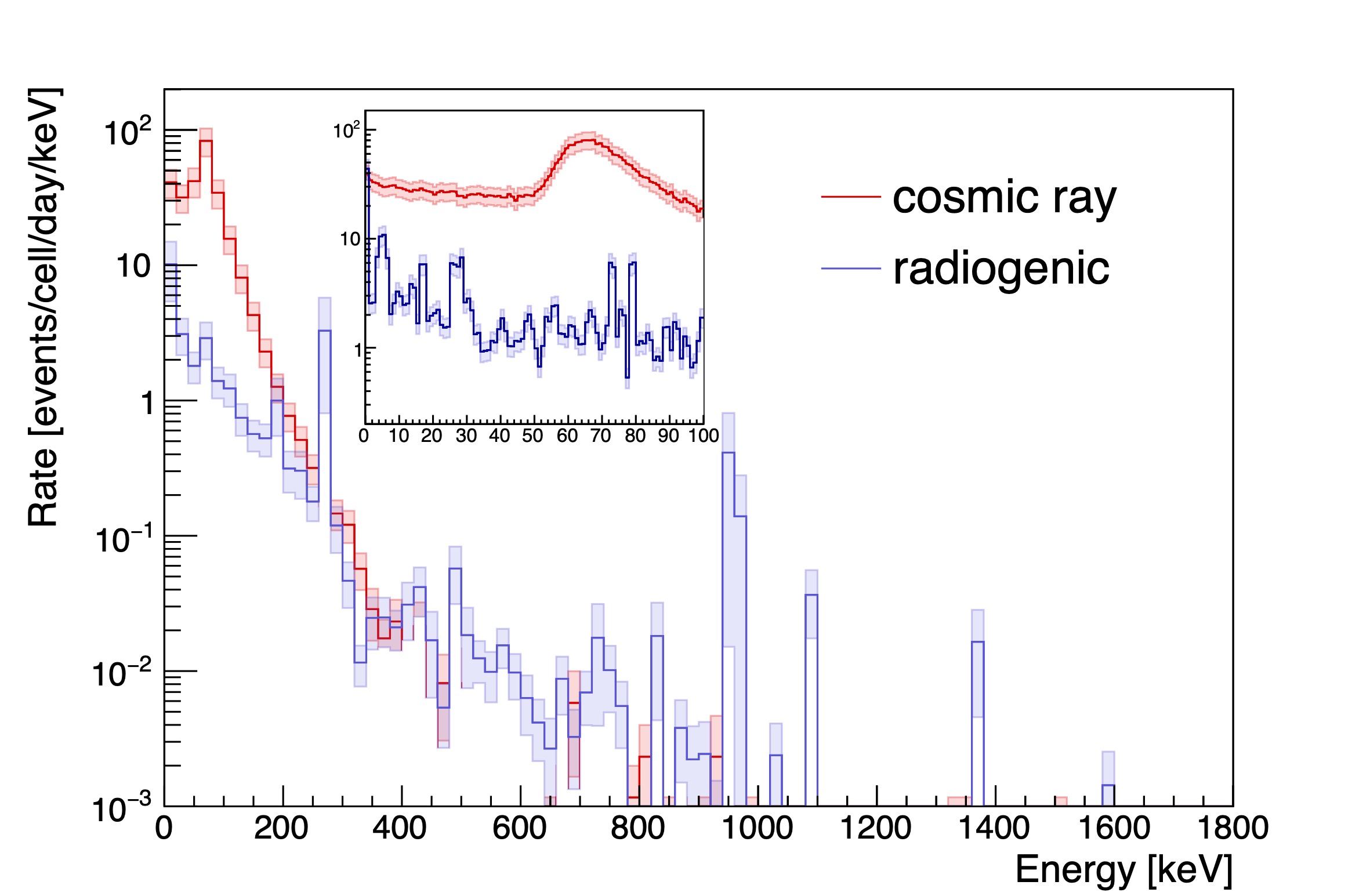}
\caption{Simulated energy spectra of background energy deposits in the $^3$He target in a QUEST-DMC experimental cell (0.315 cm$^3$, operated at 0.12 $T_c$ and saturated vapour pressure). The sum of simulated radiogenic backgrounds from materials surrounding the detector is shown, along with cosmic ray generated events (assuming operation on the Earth's surface). Shaded error bands show the sum of statistical errors and systematic errors (on the activity and flux normalisations). The inset plot shows the spectra in the 0-100 keV energy range.}\label{fig:bgspec}
\end{figure}

In order to build up a picture of the full energy deposited in an experimental cell a detailed model of the surrounding materials is required. This can be made using the GEANT4 software \cite{G4_2006, G4_2016} which simulates interactions of particles with matter across a wide range of energies, tracking their interactions and energy deposits. As described in Ref. \cite{quest_epjc} a detailed model of the QUEST-DMC detector and cryostat volumes has been constructed and for each volume $10^5$-$10^{10}$ primary decays are simulated per isotope, depending on distance from the cell. The resulting energy in the cell is recorded and normalised using screening or previous results from the SNOLAB radiopurity database \cite{radiopurity}. Energy spectra for all isotopes are summed to find the total radiogenic background expected.

Since the experiment will be located above ground we also expect a significant background from cosmic ray interactions in the cell. This is simulated using GEANT4, plus the CRY library \cite{cry} as a particle generator for incident cosmic rays. The cosmic ray flux at the Earth's surface is normalised to $\rm 0.017 /cm^2/s$ \cite{cosmic_earth}, where the uncertainty arising from the measured flux is much smaller than the statistical uncertainty in simulations. Many dark matter experiments operate underground in order to minimise the cosmic ray induced background, for example a muon flux of $(4.09 \pm 0.15) \times 10^{-8} \rm /cm^2/s$ has been measured at a depth of 1.1 km in Boulby mine \cite{Boulby_muon_2003}.

Figure \ref{fig:bgspec} shows the energy spectra expected from radiogenic decays and cosmic rays in a 0.315 cm$^3$ experimental cell in the Lancaster cryostat described in Ref. \cite{quest_epjc}, operated at saturated vapour pressure and 0.12 $T_c$, where $T_c$ is the superfluid transition temperature. The inset plot shows the spectra in the energy region below 100 keV, which is most interesting for dark matter searches. The dominant contribution to radiogenic backgrounds changes across different energy ranges. In the low energy region, below 100 keV, low energy $\gamma$ rays arising from higher activity materials further from the target dominate. In these spectra individual $\gamma$ peaks cannot be distinguished due to coarse binning and statistical fluctuations in the simulations. Rare $\alpha$ emissions from materials adjacent to the cell will dominate at high energies, above 1 MeV, as they are the only particles that can deposit this amount of energy in the cell. At intermediate energies $\beta$ emissions from materials close to the cell become important. For cosmic ray backgrounds most of the energy deposits arise from secondary electrons generated in the detector materials. Cosmic muons also deposit energy which depends on path length through the cell, giving the peak seen at $\sim$ 70 keV. Table \ref{tab:fullbg} shows the average power in the bolometer cell resulting from these energy deposits, where the cosmic ray induced power is shown for both a surface experiment and one located underground e.g. at Boulby. 

\begin{table}[t!]
\captionsetup{width=\linewidth}
\begin{tabular}{lll}
\toprule
& Events/cell/day & Average power/cell [pW]\\
\midrule
Radiogenic & $490(20)$  & $1.15(9) \times 10^{-4}$  \\
Cosmic ray surface & $5220(70)$ & $6.3(1) \times 10^{-4}$ \\
Cosmic ray underground & $1.3(1) \times 10^{-2}$ & $1.5(1) \times 10^{-9}$ \\
\botrule
\end{tabular}
\caption{Expected event rates and calculated average power for energy deposits due to radiogenic decay products and cosmic rays (or secondaries) interacting with $^3$He in a 0.315 cm$^3$ bolometer cell used in the QUEST-DMC detector (operated at saturated vapour pressure and 0.12 $T_c$). This shows an example of the full heat leak expected in a typical ultra-low temperature cryostat.}
\label{tab:fullbg}
\end{table}

\section{Conclusion}\label{sec13}

The QUEST-DMC programme aims to utilise superfluid $^3$He instrumented with nanomechanical resonators as a bolometer for dark matter detection. Design and realisation of a search for rare interactions requires detailed knowledge of potential background events. These can be modelled using Monte Carlo simulations, normalised using extensive radioassay measurement campaigns. Here, the background modelling efforts for the QUEST-DMC experiment are reported, including germanium screening results for materials commonly used in ultra-low temperature cryostats and comprehensive GEANT4 simulations of both radiogenic and cosmic ray backgrounds. These simulation results have been used to select materials for the design of the QUEST-DMC experiment and evaluate the dark matter sensitivity, as reported in Ref. \citep{quest_epjc}.

Since heating in the experimental cell is of interest beyond the dark matter community, the screening results have also been converted into heat leaks per unit mass or surface area for the different materials. The expected energy spectrum, event rate and heat leak in a single QUEST-DMC cell is shown as an example for a typical ultra-low temperature cryostat. For an experiment on the Earth's surface, with no dedicated shielding, cosmic ray backgrounds are expected to dominate, particularly at low energies. However, if such an experiment is operated underground e.g. at a depth of 1.1 km in Boulby mine the cosmic ray background is reduced by $\sim 6$ orders of magnitude. Modelling of muon energy loss at small depths based on Ref. \cite{PDG_22}, validated using Ref. \cite{Barbouti_1983}, shows that cosmic muon flux is reduced by more than an order of magnitude for depths greater than 18 m in standard rock (density 2.65 g cm$^{-3}$). At greater depths the radiogenic heat leak will dominate over cosmic ray energy deposits, limiting the experimental sensitivity -- so choice of radiopure materials and construction techniques becomes critical for any underground operation.

An increasing number of cryogenic experiments rely on isolation from interactions with the environment, specifically energy deposits. Successful operation of nuclear demagnetisation cryostats at temperatures of the order $\sim 0.1$ mK depends on minimising heat leaks to below the pW level \cite{Bradley84}. Once thermal and vibrational isolation has been optimised it may be important to consider heat generated by ionising radiation from radioactivity and cosmic rays, as done in Ref. \cite{nazaretskiheat_04}. In recent years superconducting circuits and qubit technology have improved sufficiently that this is also reaching the point of being limited by energy deposits from cosmic rays or radioactive decays \cite{Cardani_2021}, which must be well understood to enable robust error correction \cite{Wilen_21}.

\backmatter

\bmhead{Acknowledgements}
This work was funded by UKRI EPSRC and STFC (Grants ST/T006773/1, EP/P024203/1, EP/W015730/1 and EP/W028417/1), as well the European Union's Horizon 2020 Research and Innovation Programme under Grant Agreement no 824109 (European Microkelvin Platform). M.D.T acknowledges financial support from the Royal Academy of Engineering (RF\textbackslash 201819\textbackslash 18\textbackslash 2),  We thank Boulby Underground Laboratory for radioassaying the materials.

\bibliographystyle{sn-aps}
\bibliography{qfs-bib}

\end{document}